\documentclass[oneside, 11pt, parskip=full]{scrartcl}

\pdfoutput=1
\usepackage{lscape}
\usepackage[super,square,comma,sort,compress]{natbib}
\usepackage{graphicx}
\usepackage{amsmath} 

\usepackage{tabularx}

\usepackage[a4paper,
top=2.5cm,
bottom=1.5cm,
includefoot,
left=2.5cm,
right=2.5cm]{geometry}

\usepackage{caption} 
\DeclareCaptionStyle{mystyle}%
[margin=5mm,justification=centering]%
{font=small,margin={0mm,0mm}} 
\begin{document}
	
	\begin{center}
		
		\textbf{\LARGE{Photophysics of two-dimensional perovskites \textendash\ learning from metal halide substitution}}

		{\Large Simon Kahmann,$^{a,b}$ Herman Duim,$^{a}$ Hong-Hua Fang,$^{a,c}$ Mateusz Dyksik,$^{d,e}$, Sampson Adjokatse,$^{a}$ Martha Rivera Medina,$^{a}$ Matteo Pitaro,$^{a}$ Paulina Plochocka,$^{d,e}$ Maria~A.\ Loi$^{a}$}
		
		\textit{$^{a}$ Photophysics and OptoElectronics, Zernike Institute for Advanced Materials, University of Groningen, Nijenborgh 4, 9747 AG, Groningen, The Netherlands\\
		$^{b}$ Current address: Cavendish Laboratory, University of Cambridge, JJ Thomson Avenue, CB30HE Cambridge, United Kingdom \\
		$^{c}$ Current address: State Key Laboratory of Precision Measurement Technology and Instrument, Department of Precision Instrument, Tsinghua University, Beijing 100084, China\\
		$^{d}$ Laboratoire National des Champs Magn\'{e}tiques Intenses, UPR 3228, CNRS-UGA-UPS-INSA, Grenoble and Toulouse, France\\
		$^{e}$ Department of Experimental Physics, Faculty of Fundamental Problems of Technology, Wroclaw University of Science and Technology, Wroclaw, Poland\\
		}
				
	\end{center}

\begin{abstract}

		Whereas their photophysics exhibits an intricate interplay of carriers with the lattice, most reports have so far relied on single compound studies. With the exception of variations of the organic spacer cations, the effect of constituent substitution on the photophysics and the nature of emitting species, in particular, has remained largely under-explored.\\
		Here PEA$_2$PbBr$_4$, PEA$_2$PbI$_4$, and PEA$_2$SnI$_4$ are studied through a variety of optical spectroscopy techniques to reveal a complex set of excitonic transitions at low temperature. We attribute the emergence of weak high energy features to a vibronic progression breaking Kasha's rule and highlight that the responsible phonons cannot be accessed through simple Raman spectroscopy. Bright peaks at lower energy are due to two distinct excitons, of which the upper is a convolution of a bright exciton and a localised state, whereas the lower is attributed to shallow defects.\\
		Our study offers deeper insights into the photophysics of two-dimensional perovskites through compositional substitution and highlights critical limits to the communities' current understanding of the photophysics of these compounds.
				
\end{abstract}

\textbf{INTRODUCTION}

What is the impact of metal and halide substitution in two-dimensional perovskites? How does it affect the optical properties of these layered compounds and can we attain deeper insights into the materials'  photophysics through comparison? Two-dimensional perovskites consist of metal halide (BX$_6$) octahedra sandwiched between long organic spacer molecules (A) that impose the low-dimensional electronic character through both a quantum and a dielectric confinement of charge carriers into the inorganic layer. As a result of this confinement, the exciton binding energy in these systems exceeds 100\,meV\cite{Ishihara1989} leading to a bright luminescence that renders 2D perovskites hot contenders for opto-electronic applications. This also demands for a thorough understanding of these compounds' photophysics, as formulated in the preceding questions.\\
Reflecting the focus of three-dimensional metal halide perovskite research (HaPs), most studied compositions in the 2D field have so far been based on Pb as the B-site metal cation. Most commonly, either iodide or bromide have been used as halide anion on the X-site, to obtain, for example, green or blue luminescence.\cite{Blancon2020}  As suggested by these different colours, halide variation strongly impacts the materials' band gap.\cite{Marchenko2020} A larger variety of organic moieties has been employed as the A-site spacer cation, of which the prototypical molecules are butylammonium (BA$^+$) or phenylethylammonium (PEA$^+$). The effect of the A-site cation on the optical properties is generally smaller; besides the straightforward change of the confinement through variations in layer separation and permittivity, the molecules also impact the position and the degree of distortion of the inorganic species.\cite{Du2017a,Katan2019,Marchenko2020,Dyksik2020}

Given the large binding energy, these compounds are excellent systems to study exciton physics. Low temperature studies offer complex data on both absorption and light emission characteristics, which are hotly debated in the field. Especially phenylethylammonium lead iodide (PEA$_2$PbI$_4$) has recently become a working horse with reports claiming biexciton emission,\cite{Fang2020} large exchange interaction splitting,\cite{Do2020} defect-induced luminescence,\cite{Kahmann2020} pronounced hot exciton luminescence,\cite{Straus2016} and the formation of exciton polarons.\cite{Thouin2018} The nature and impact of carrier-phonon interactions, as considered in the latter reports, are also of general importance to the wider field of HaP research. Many unusual phenomena, such as long carrier lifetimes, broad Raman spectra, or broad luminescence bands are attributed to a peculiar behaviour of phonons and their interaction with charge carriers.\\
As mentioned above, with few exceptions including different spacer cations or tin incorporation,\cite{Straus2019,Straus2020,Folpini2020,Tekelenburg2021} photophysical studies have so far relied predominantly on lead iodide-based compounds with either BA or PEA as the A-site cation. Studied phenomena have thus typically been considered irrespective of related compounds in the broader field of the compositional space. In other words, the impact of the metal and halide constituents has so far remained critically under-explored.\\
Here we aim to fill this gap of knowledge and we provide and in-depth study of the photophysical properties of a set of two-dimensional HaPs based on PEA as the A-site cation. Additional to the standard PEA$_2$PbI$_4$ compound, we introduce bromide as alternative halide on the X-site and tin as substituent of lead on the metal B-site. We focus our study on the nature of the excitonic properties around the band edge and consider our results in the framework of current discussions in the field. We find that all compounds give rise to an intricate sub-structure of PL peaks at low temperature. High energy luminescence directly mirrors absorption features and is consequently linked to the recombination of hot excitons that violate Kasha's rule. Our variational approach allows us to show that a simple one-to-one correspondence between A-site vibrational modes detected in Raman spectroscopy, and the spacing of this vibronic progression is invalid. Bright luminescence at lower energy is furthermore attributed to diffusion limited recombination from crystal imperfections and we employ magneto-PL studies to show evidence of an anomalous exciton bright-dark splitting with a simultaneous localisation of the 1\textit{s} exciton.

\textbf{RESULTS}

\begin{figure}[]
	\centering
	\includegraphics[width=.5\linewidth]{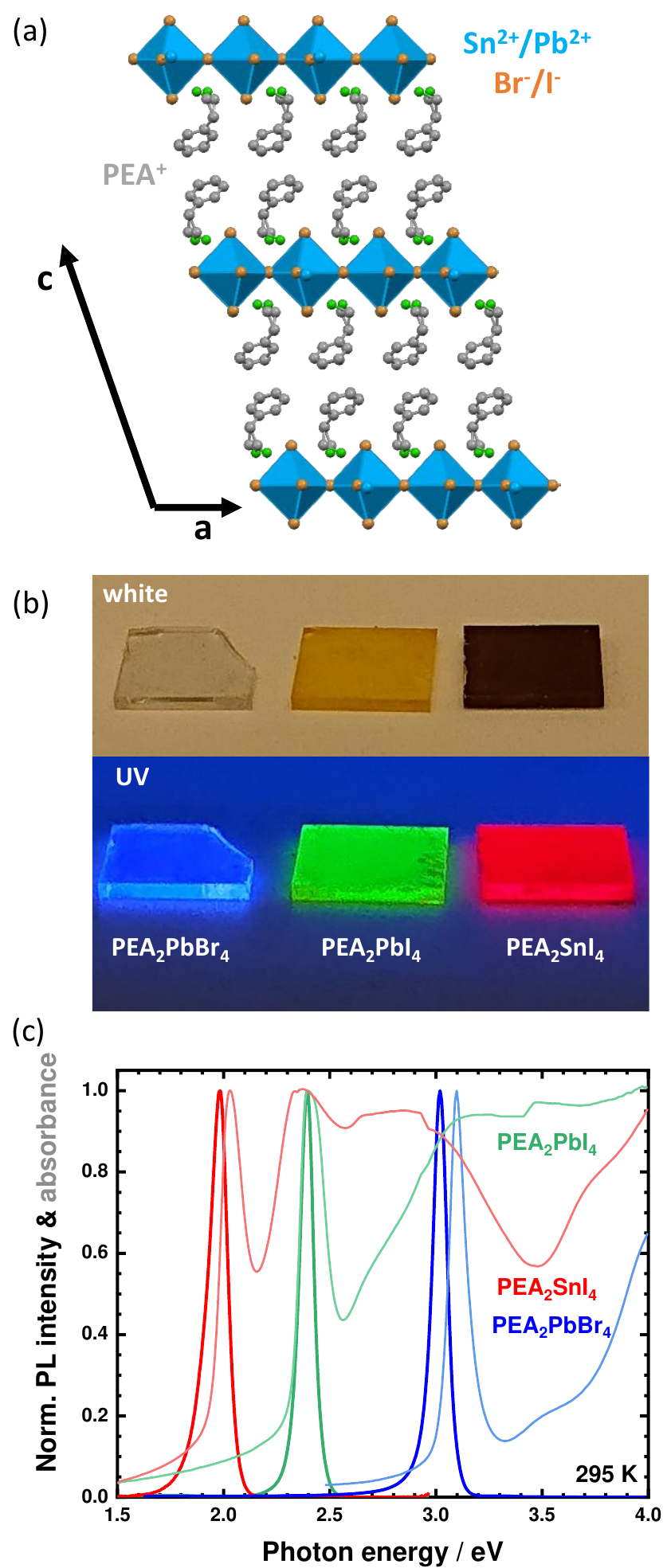}
	\caption{(a) PEA$_2$BX$_4$ crystal structure incorporating alternating layers of metal halide cages and PEA$^+$ spacer cations. (b) Photographs taken under white light (top) and UV-illumination (bottom). (c) Absorbance (lighter colours) and PL spectra (darker) at room temperature exhibit a strong excitonic resonance with a small Stokes shift. Spectra are normalised to the exciton transition. $\lambda_{ex}$=4.6\,eV for PEA$_2$PbBr$_4$, else 3.1\,eV.}
	\label{fig1}
\end{figure}

We prepared a set of compounds based on phenylethylammonium (PEA) as the A-site spacer cation. Additional to the prototypical lead iodide-based compound, we considered the variants based on lead bromide and tin iodide. The layered structure of the films, highlighted in Figure~\ref{fig1} (a), forms in out-of plane direction, meaning that the organic and inorganic slabs run parallel to the sample surface, as confirmed by X-ray diffraction (Supplementary Note~1). As expected, the compounds possess distinctly different appearances both under white light and in particular under UV illumination, as shown in Figure~\ref{fig1} (b). Analogous to their three-dimensional counterparts, the valence band maximum of two-dimensional perovskites is determined by the metal's outermost \textit{s}-orbital and the halide's \textit{p}-orbital, whereas the conduction band is determined by a combination of the metal and halide \textit{p}-orbitals as well as the metal \textit{p} and halide \textit{s}-orbital.\cite{Fabini2020,Zhang2020a} Variation of the constituent metal and halide species thus exhibits the analogous effect on the band gap (E$_{gap}$): a widening of the gap for smaller halides and a narrowing when substituting lead with tin.\cite{Goesten2018} This translates into the shifted absorption onset and PL peak energy shown in Figure~\ref{fig1} (c).\\
Smaller contributions to band gap energy variations arise from different degrees of distortion in the inorganic layer, but, as we discuss in Supplementary Note~1, the here employed compounds exhibit largely similar degrees of distortion. This is highly beneficial, as it allows for drawing conclusions based on the chemical nature of the substituted constituents without the need to assess secondary impacts from changes in the crystal structure.\\
All studied compounds exhibit a narrow absorption band due to the excitonic transition and strongly increasing absorption of continuum transitions towards high energy. The luminescence is governed by a narrow emission and a small Stokes shift, whose shape and position are independent of the incident laser power (Fig.~S1). The PL intensity furthermore scales linearly with the incident power, all underlining the excitonic origin of the luminescence.\\
Before focusing on the main issues of this report, we would like to note that in a stark contrast to the 3D compounds,\cite{Parrott2018,Kahmann2019a} the time resolved PL of the lead-iodide compound decays significantly faster than that of the tin-based variant. The latter gives rise to a persistent PL tail ranging toward the microsecond time scale (Fig.~S2), indicating the formation of long-lived states upon photoexcitation.

\begin{figure}[]
	\centering
	\includegraphics[width=.5\linewidth]{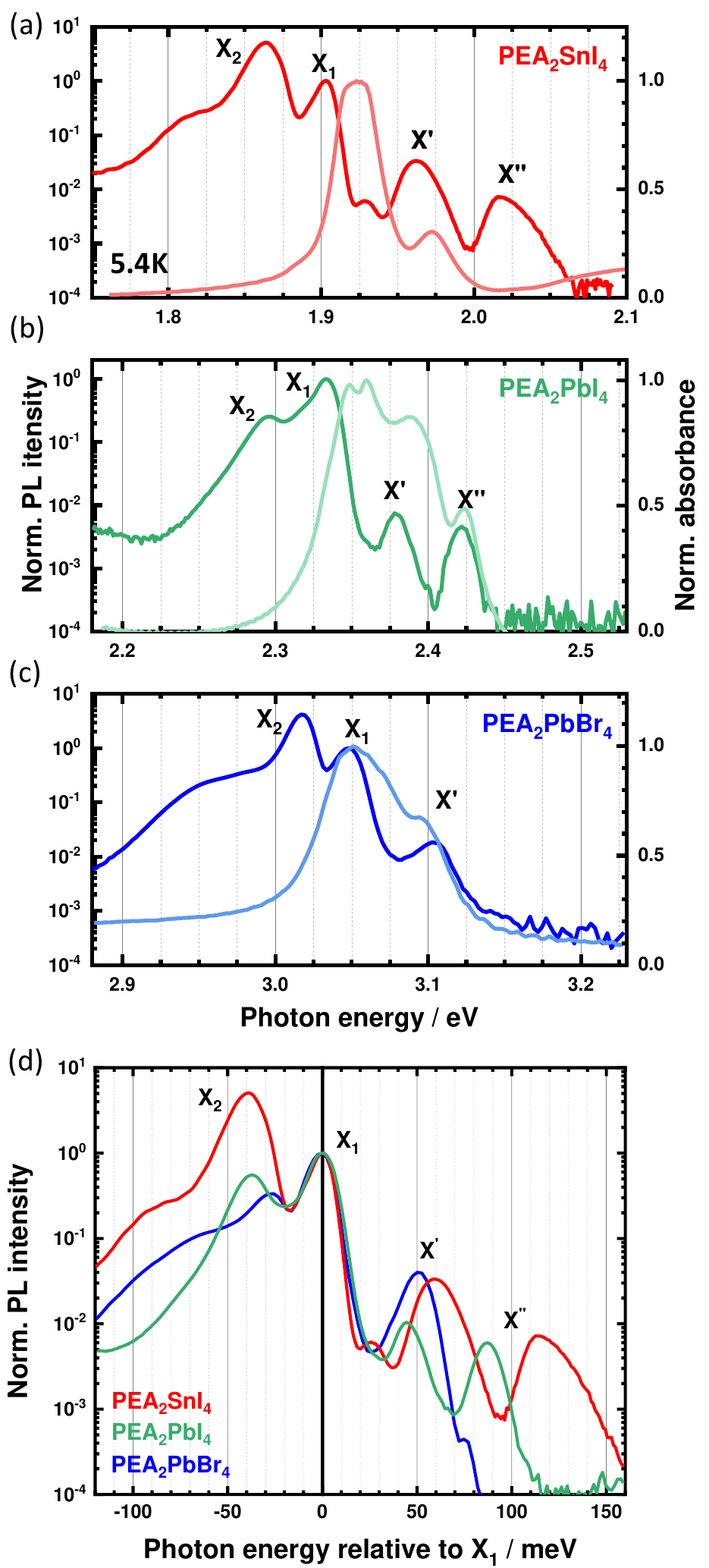}
	\caption{(a)-(c) PL spectra of the three compounds at 5.4\,K normalised to the upper main peak, X$_1$ with corresponding normalised absorbance spectra. Note the logarithmic intensity axis of the luminescence. (d) PL spectra aligned with respect to X$_1$.}
	\label{fig2}
\end{figure}

At low temperature, the narrow band edge emission splits into a rich substructure, shown in Figure~\ref{fig2}. The intensity of these lines varies over several orders of magnitude and each spectrum is centred around two pronounced emission lines, X$_1$ and X$_2$ with one or two weak peaks at higher energy and broader contributions at energies below. Table~\ref{tab1} summarises the positions.\\
To understand the origin of the PL, it is expedient to consider the luminescence in context of the samples' absorption. Figure~\ref{fig2} displays the corresponding data in panels (a) to (c). In contrast to the room temperature spectra in Figure~\ref{fig1} (c), also the absorbance develops a distinct substructure upon cooling. Instead of a mere narrowing of the 1$s$ exciton line, several absorption peaks emerge on the high energy side (note that PEA$_2$PbI$_4$ is partially saturated around 2.35\,eV). The X$_1$ emission exhibits a small Stokes shift relative to the lowest absorption peak for all three cases and we thus conclude that X$_1$ is due to the recombination of the ground state (GS) exciton.\\
Moreover, all three samples give rise to at least one distinct absorption peak on the high energy side that coincides with either X$^\prime$ or X$^{\prime\prime}$ and a close look reveals that all observable PL lines at high energy coincide with distinct absorbance features (also consider \textbf{Fig.~S4}). Importantly, the small Stokes shift and the overlap of PL and absorbance can lead to significant reabsorption of the luminescence affecting magnitude and shape of the observed peaks. For example, Figure~\ref{fig2} (a) reveals the above mentioned small peak 25\,meV above X$_1$ for PEA$_2$SnI$_4$ to lie at the maximum absorption, which will diminish the observable intensity for this peak. A comparison with the lead-based samples suggests that also these exhibit a shoulder between X$_1$ and X$^\prime$, which remains unresolved due to spectral broadening.

Given their similarities around the band edge, we replotted the PL spectra in Figure~\ref{fig2} (d). The high energy peaks of all three samples lie within a range of 150\,meV from the main emission peak X$_1$. For PEA$_2$PbI$_4$ and PEA$_2$SnI$_4$, the two high energy peaks are virtually equidistant at 45\,meV and 87\,meV and 60\,mev and 113\,meV for Pb and Sn, respectively (consider Table~\ref{tab1}). Both peaks of the latter are furthermore much broader than for the lead-based compounds and clearly asymmetric. Even weaker contributions can furthermore be resolved in the background, for example for PEA$_2$SnI$_4$ around 25\,meV \textendash\ underlining the richness of the substructure. PEA$_2$PbBr$_4$, in contrast, only exhibits one distinct peak approximately 51\,meV above X$_1$, which we assume is due to the diminishing detector sensitivity at even higher energy ($>$3.2\,eV).

\begin{table}[t]
	\centering
	\caption{PL peak position at 5.2\,K given in eV and their distance from X$_1$ in meV in parentheses}
	\begin{tabular}{lllll}
		\hline
		Compound & X$_2$ & X$_1$ & X' & X'' \\
		\hline
		PEA$_2$PbBr$_4$ & 3.012 (-31) & 3.043 (-) & 3.095 (51) & - \\
		PEA$_2$PbI$_4$ & 2.298 (-38) & 2.334 (-)	& 2.379 (45) & 2.422 (87) \\
		PEA$_2$SnI$_4$ & 1.863 (-39) 	& 1.903 (-)	& 1.962 (60) & 2.017 (113) \\
		\hline
	\end{tabular}
	\label{tab1}
\end{table}

\begin{figure}[]
	\centering
	\includegraphics[width=1\linewidth]{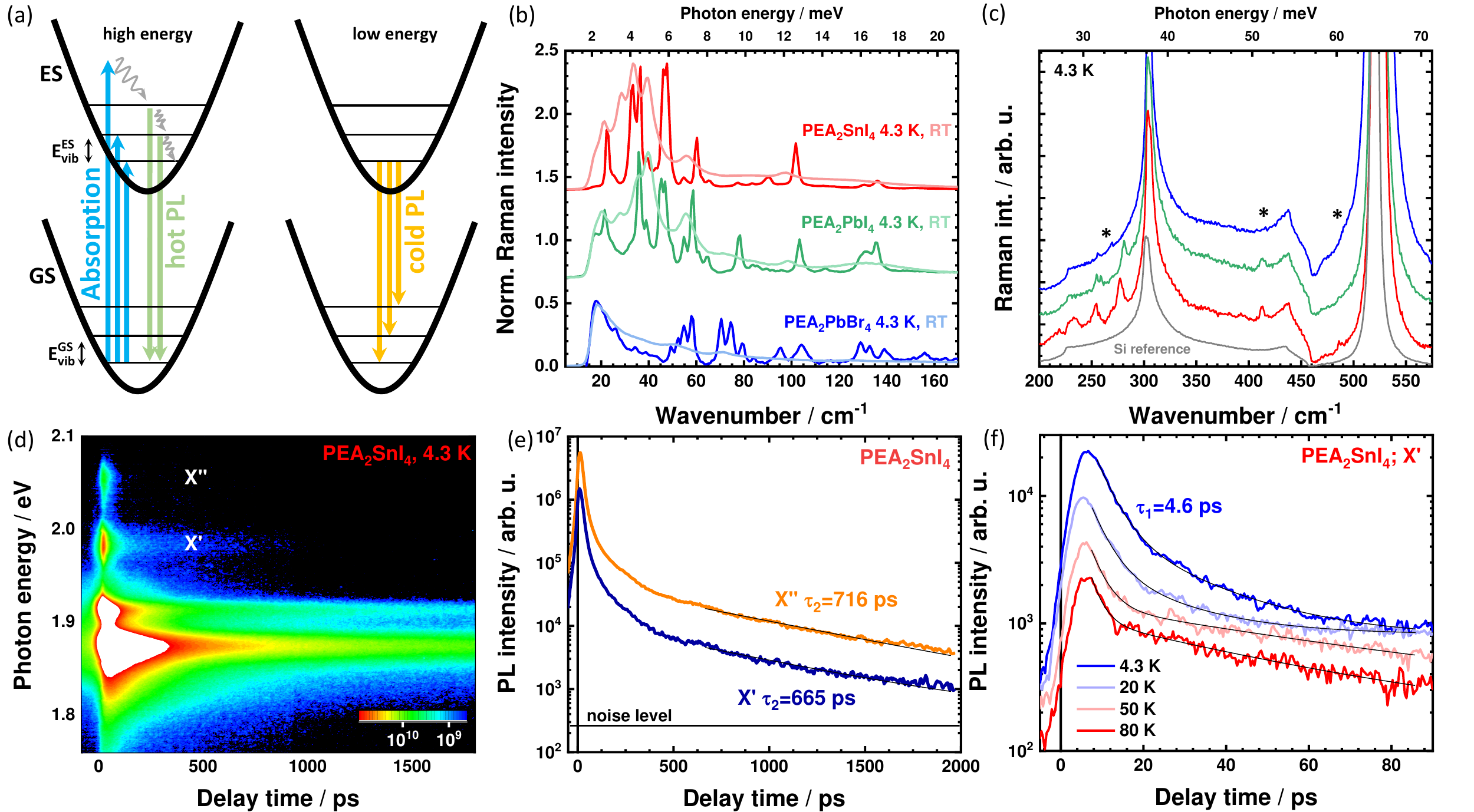}
	\caption{(a) Relevant transitions of the absorbance, hot, and cold luminescence sketched in a Frank-Condon framework. (b) Non-resonant Raman spectra at room temperature and at 4.3\,K in the low energy region. (c) Region corresponding to the spacing of the high energy PL lines, as indicated by asterisks. The pronounced background stems from the silicon substrate shown in grey. (d) Streak camera plot of the PEA$_2$SnI$_4$ PL at 4.3\,K and (e) corresponding extracted decay curves of the hot luminescence. (f) Short delay decay dynamics upon temperature variation.}
	\label{fig4}
\end{figure}

The symmetry between the absorption and the high energy luminescence shows that the respective peaks originate from the same transitions. Considering the similar spacing of the high energy features, the structure can be understood as a Frank-Condon series. The observed absorption peaks stem from transitions originating from the electronic ground state ending on a vibronic progression of the excited state (ES), as depicted in Figure~\ref{fig4} (a). The corresponding PL lines analogously stem from the inverse process. In other words, recombination occurs from excited vibrational sub-levels, so-called hot excitons, in express violation of Kasha's rule. The observed peak separation thus results from the coupling of excitons to vibrational modes of energy E$_{vib}$.\\
We note that one might consider these features to be due to a Rydberg series, but both their spacing as well as their common shift in magnetic fields rule out this possibility, as recently reported for the iodide-based compounds.\cite{Dyksik2020,Urban2020}

Such a vibronic progression was proposed for PEA$_2$PbI$_4$ before\cite{Straus2016} and the dominant mode coupling to the exciton was said to stem from the spacer molecule on the A-site; a claim subsequently backed up by considering several A$_2$PbI$_4$ compounds.\cite{Straus2020} In this light the large differences in spacing between the current samples surprise, given that the compounds are all based on PEA on the A-site. Certainly, the interplay of the molecules with different inorganic species will affect the energy of relevant vibrational modes, but the small differences in crystal structure and distortion (Supplementary Note~1), render a 15\,meV difference between PEA$_2$SnI$_4$ and PEA$_2$PbI$_4$ unexpectedly large.\\
We therefore monitor the vibrational structure of the compounds through non-resonant Raman scattering ($\lambda$=785\,nm; 1.58\,eV; for further discussion consider Supplementary Note~2). Pronounced Raman signals in HaPs typically stem from the inorganic components found in the region below 80\,cm$^{-1}$ (10\,meV).\cite{Niemann2016} These features are evident in Figure~\ref{fig4} (b), which compares the low energy region at room temperature and at 4.3\,K. Temperature reduction allows for distinguishing a rich subset of peaks that appear as broad features at elevated temperature. The strong response obtained from PEA$_2$PbI$_4$ and PEA$_2$SnI$_4$ facilitates the direct comparison to extract the impact of metal variation on the energy of vibrational modes. The shift of dominant peaks in the low energy region is small (approx.\ 1\,meV). In contrast, a stronger shift (and smaller response) manifests for PEA$_2$PbBr$_4$, which is in accordance with previous data.\cite{Dhanabalan2020}\\
The spectral region directly corresponding to the separation of the hot luminescence is shown in Figure~\ref{fig4} (c). The scattering intensity of modes in this region is several orders of magnitude smaller and the spectra are dominated by well known features from the silicon substrates.\cite{Uchinokura1974} Nonetheless, several signals from the perovskites can be identified and the respective regions of interest are indicated. Phonon modes in this region can in part be attributed to modes stemming from the organic moiety.\cite{Straus2016,Ni2017a,Urban2020} Given its comparable energy, the phonon pair around 32\,meV was previously said to determine the spacing of PEA$_2$PbI$_4$ by coupling to the excitons.\cite{Urban2020} Considering the current results, such a direct correspondence should be made with care: not only is this energy significantly smaller than the PL spacing of 45\,meV, but these Raman modes appear similarly also for PEA$_2$SnI$_4$ and for PEA$_2$PbBr$_4$, whose PL peaks are separated by approximately 60 and 50\,meV. Conversely, a minor feature around 60\,meV in the Raman spectrum of PEA$_2$SnI$_4$, which one might identify as the coupling mode, can also be observed for PEA$_2$PbI$_4$.\\
Obviously, there is a discrepancy between the differences in the spacing of PL/absorption peaks and the uniformity of the Raman spectra in the relevant spectral region. There is no direct correspondence between the observable Raman modes and the separation of features in PL/absorption. This means that despite clear evidence from magneto-absorption\cite{Dyksik2020} and spectral symmetry for a vibronic progression to determine the PL and absorption lines, the mechanism of exciton phonon coupling remains elusive. Straightforward attempts linking the energy of Raman modes to the spacing are inadequate and the lack of direct correspondence requires further investigations.\\
A possible explanation might be strong differences between the vibrational properties of the ground state ($E^{GS}_{vib}$ probed in Raman spectroscopy and the excited state ($E^{ES}_{vib}$), which determines the spacing in hot PL and absorbance (Fig.~\ref{fig4} (a)). So far reported data on the ES have shown great similarities with the GS,\cite{Ni2017a} but the current data clearly warrant further study.\\

Another consequence of the vibronic progression is that hot exciton recombination competes with the relaxation towards the lowest excited state, as sketched in Figure~\ref{fig4} (a). Additional to a low intensity of X$^\prime$ and X$^{\prime\prime}$, this should result in a short lifetime of the hot luminescence. Not surprisingly, reported lifetimes for X$^\prime$ and X$^{\prime\prime}$ lie well below 10\,ps.\cite{Straus2016,Straus2020} In stark contrast, Figure~\ref{fig4} (d) shows a streak camera image (PEA$_2$SnI$_4$; 4.3\,K) that displays a long-lived tail for X$^\prime$. The decay curves prove this to be true for both hot peaks (Fig.~\ref{fig4} (e)). The initial rapid decay (resolution limited around $\tau_1$=4.5\,ps) reduces the signal by two orders of magnitude, but is followed by a second, much slower process with a lifetime of around 700\,ps. The presence of this two-stage decay is similarly observed at elevated temperature, where the fast initial decay becomes less dominant (Figure~\ref{fig4} (c)), and it is analogously present for the two lead-based compounds (Fig.~S5).\\
To the best of our knowledge, this long-lived process has so far not been observed. Based on its presence in all three compounds, we infer that it is a general characteristic of 2D HaPs. The slow decay of X$^\prime$ and X$^{\prime\prime}$ luminescence conflicts with the rapid relaxation of excitons towards the lowest vibrational level of the ES. As possible explanations, Auger processes or exciton-exciton annihilation could promote relaxed excitons into higher states and thereby repopulate excited vibrational levels. In this case, we would expect a strong dependence of the PL lifetime on the excitation intensity, but we found $\tau_2$ to remain unaffected over the excitation range accessible to us. A more promising explanation might be the retarded recombination of free charge carriers. Given that the employed high energy excitation initially generates free charge carriers, it is possible for a small, but finite, population of charge carriers to remain unbound and to only form excitons at a later stage. The long-lived PL tail of X$^\prime$ and X$^{\prime\prime}$ would thus be dictated by the lifetime of free charge carriers and not by the exciton relaxation. This hypothesis should be verifiable through optical or terahertz pump-probe spectroscopy.

We now turn to the low energy side of the spectra, to what can be termed the cold luminescence from relaxed states. As evident from Figure~\ref{fig2}, all three samples exhibit a second pronounced PL peak (X$_2$) and further smaller contributions below X$_1$ at low temperature. We focus on signals shifted by less than approximately 50\,meV, noting that features at lower energy involve either shallow or deep defects.\cite{Kahmann2020}\\
Based on the above discussion and on low temperature spectra alone, one might assume the structure of low energy luminescence to merely be due to transitions towards higher vibrational levels of the GS, as shown in Figure~\ref{fig4} (a) on the right. However, the analysis of temperature dependent and time resolved PL shows that this is not the case.

\begin{figure}[]
	\centering
	\includegraphics[width=1\linewidth]{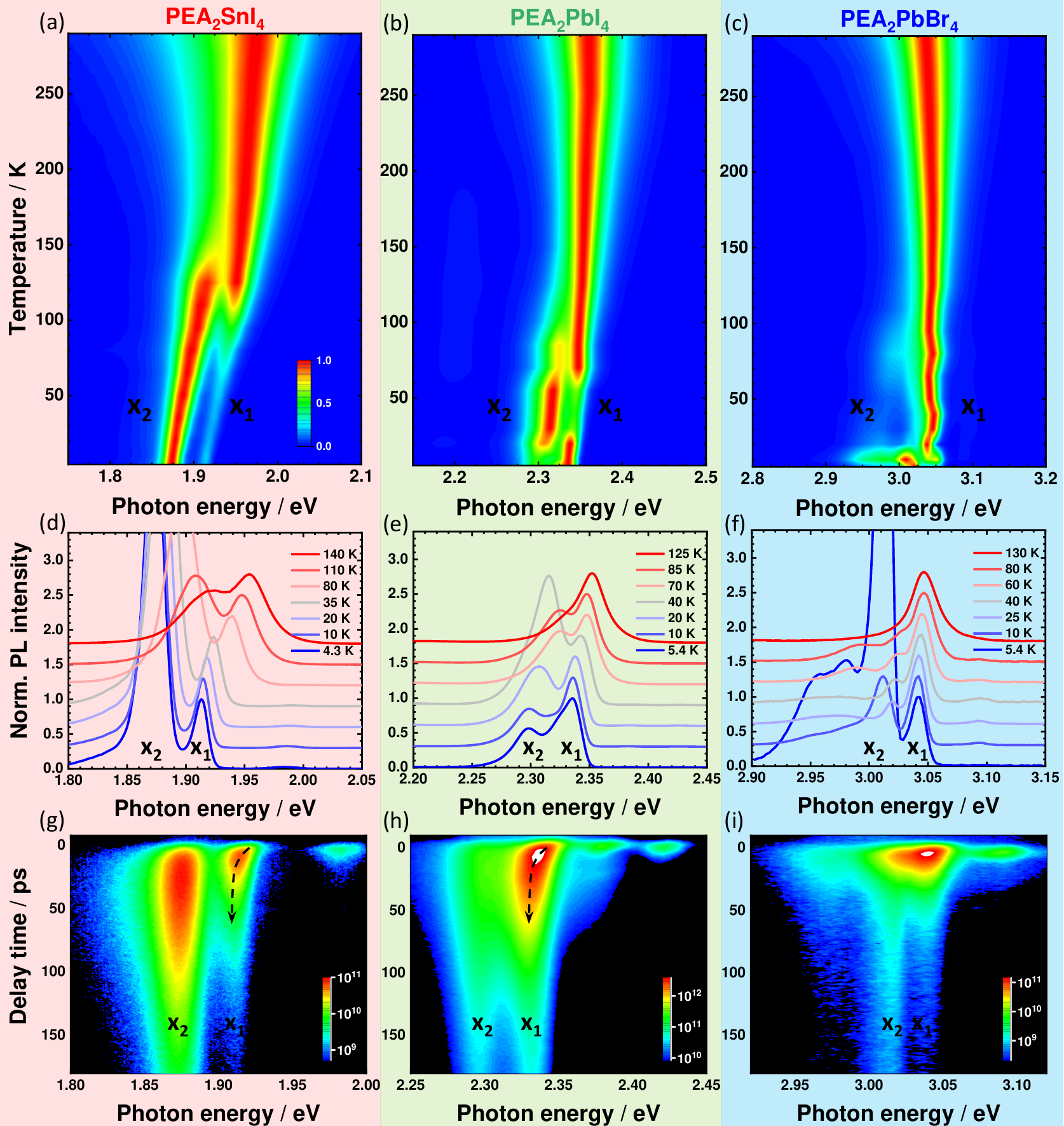}
	\caption{Temperature dependent PL of the three materials normalised to a global maximum in (a), (b), and (c) and to X$_1$ in (d), (e), and (f). Spectra in (d)-(f) are vertically offset. Temperature variation has a profound effect on the luminescence in each case and the trends cannot be explained through a vibronic progression. Early time transient PL in (g), (h), and (f) highlights a delayed maximum for X$_2$ and a time-dependent red-shift of X$_1$ at early delay, indicated by the dashed arrow.}
	\label{fig5}
\end{figure}

We discuss the temperature dependent changes in more detail in Supplementary Note~3 (absorbance, steady state, and transient PL; Fig.~S10-S14) and focus here on the following aspect: the emergence of additional peaks does not merely follow from a narrowing of spectral features, but by drastic changes in PL intensity and intensity ratios, as shown in Figure~\ref{fig5} (a) to (f). For PEA$_2$SnI$_4$ the trend is relatively straightforward: Below 140\,K the emission splits into two strong peaks, X$_1$ and X$_2$ (Fig.~\ref{fig5} (a)), of which X$_2$ becomes increasingly more pronounced upon cooling (Fig.~\ref{fig5} (d); Fig.~S12).\\
The lead-based compounds show more complex behaviours. PEA$_2$PbI$_4$ also exhibits a splitting of PL peaks, but whereas X$_2$ and a third state between X$_1$ and X$_2$ are dominant from 70\,K on, X$_1$ is again the strongest below 20\,K. Similarly, the bromide-based variant exhibits a complex trend of several weaker sub-peaks around 3\,eV over the intermediate temperature region, but it is only around 20\,K and below, where X$_2$ becomes dominant.\\
The transient PL shows strikingly different dynamics for the two main peaks at low temperature, as shown in Figure~\ref{fig5} (g) to (i). X$_1$ exhibits a fast initial intensity reduction that is accompanied by a red-shift \textendash\ particularly pronounced for PEA$_2$SnI$_4$ and PEA$_2$PbI$_4$. X$_2$, on the other hand, shows a less pronounced spectral shift, but a clear delayed onset of the intensity. The extracted transients are discussed in Supplementary Note~4 (Fig.~S15-S17), where we also employ two-photon excitation to show the red-shift to not be merely through re-absorption effects. Note that despite our focus on the early stage of the decay, the overall luminescence lifetime lies in the range of microseconds (Fig.~S2, S3).

\begin{figure}[]
	\centering
	\includegraphics[width=\linewidth]{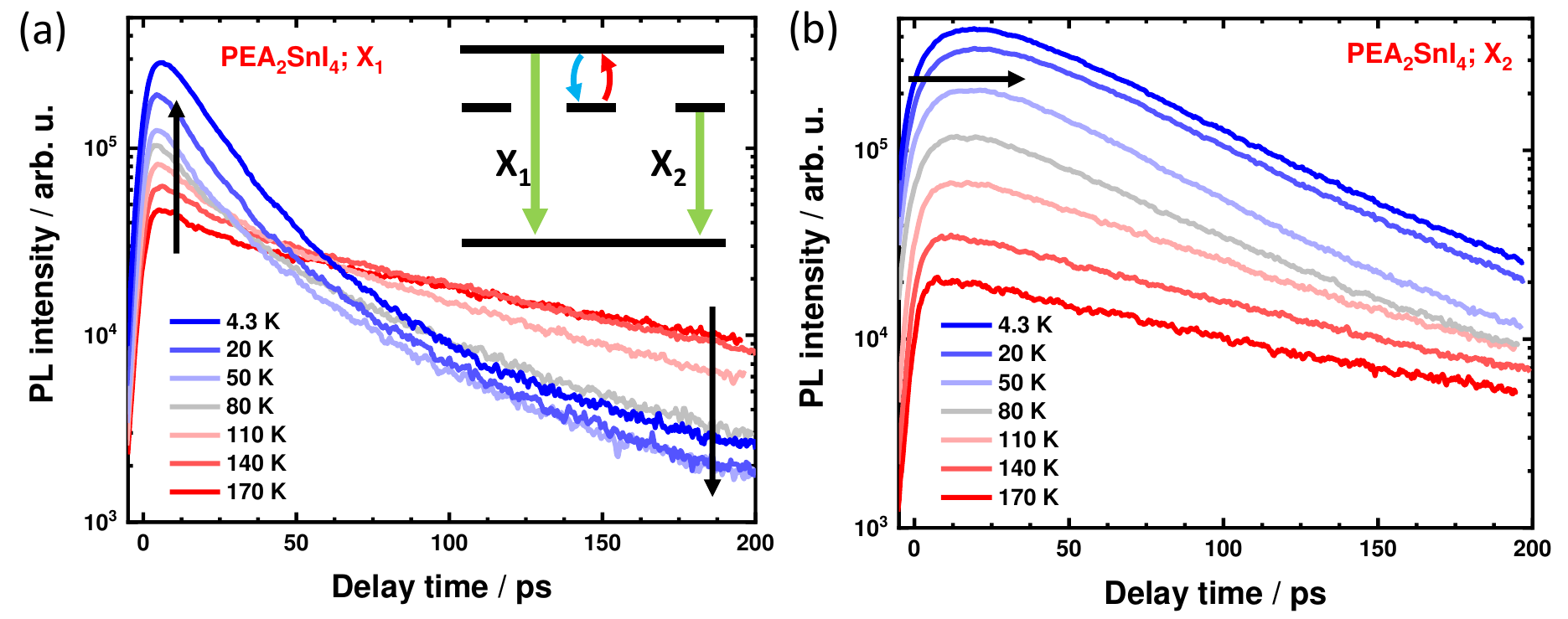}
	\caption{Temperature dependent PL for X$_1$ (a) and X$_2$ (b) of PEA$_2$SnI$_4$ at short delays. Cooling increases the overall initial intensity fo both peaks, but a fast decay component becomes increasingly dominant for X$_1$ at low temperature and a delayed maximum forms for X$_2$. The interplay is explained through a two-state model with a diffusion step, as shown in the inset of (a).}
	\label{fig6}
\end{figure}

The interplay of the two states can best be illustrated when considering the relatively simple case of PEA$_2$SnI$_4$. Figure~\ref{fig6} (a), (b) shows the temperature dependent dynamics of X$_1$ and X$_2$ at short delay time. Cooling increases the prompt PL intensity in both cases, probably due to the suppression of phonon-assisted non-radiative decay channels. However, whereas X$_2$ retains this higher intensity over time, X$_1$ shows an increasingly pronounced rapid initial decay component with $\tau$=12-15\,ps. At 170\,K the decay is almost monoexponential, but becomes clearly biexponential below. This trend is similarly observable when considering longer delays up to 2 nanoseconds, as shown in (Fig.~S12). Simultaneously, X$_2$ exhibits a monoexponential decay at 170\,K with an increasingly pronounced delayed maximum upon cooling (black arrow in Figure~\ref{fig6} (b)).\\
The pronounced spectral changes upon temperature variation along with the stark differences in early decay dynamics exclude a vibronic progression as the dominant mechanism behind X$_1$ and X$_2$. These transitions would otherwise follow identical dynamics, as they originate from the same state (Fig.~\ref{fig4} (a)) and their relative intensity should be independent of the temperature. Moreover, as shown in Figure~S6, PbI$_2$ exhibits a similar behaviour, which rules out an origin solely reliant on the organic moiety.

Historically, the X$_2$ luminescence in these layered systems has often been attributed to the recombination of biexcitons to form the so-called M-band.\cite{Kato2003,Tanaka2005,Goto2006,Fang2020} This was commonly based on a superlinear dependence of the PL intensity on the excitation fluence. Although biexcitonic emission seems already to be excluded from above findings, we nonetheless consider the impact of excitation intensity variation to clarify this point. As we discuss in Supplementary Note~5 (Fig.~S18-S20), the power dependent measurements on our films reveal two important insights: Firstly, the intensity of X$_1$ scales linearly with the excitation flux and correlates with the intensity of X$^{\prime}$ and X$^{\prime\prime}$. This underlines their shared excitonic origin, as discussed in the first part of this manuscript. Secondly, there is no uniform dependence of X$_2$ on the excitation power: PEA$_2$SnI$_4$ is strongly superlinear, PEA$_2$PbBr$_4$ sublinear, and PEA$_2$PbI$_4$ becomes superlinear at high power. The spectra are additionally subject to irreversible changes upon high intensity exposure. In particular PEA$_2$SnI$_4$ exhibits a much more pronounced X$_2$ emission after illumination.\\
We thus conclude that at the low intensities employed in our work biexcitons play no role in the emission. This is also in accordance with our study on single crystals of lead iodide based compounds.\cite{Tekelenburg2021}\\
In contrast, when considering X$_2$ relative to the absorption, as in Figure~\ref{fig2}, one finds that the emission coincides with a tail below the first absorption maximum in all three cases. Above discussed data plotted on a logarithmic axis (Fig.~S4) show that for some cases this tail exhibits a clearly resolvable absorption peak, which is in accordance with previous reports.\cite{Ni2017a,Neutzner2018,Do2020,Li2019h} There is thus a second distinct and optically allowed electronic state below the 1$s$ exciton.\\
Combined with the delayed onset dynamics and the temperature dependence of the luminescence intensity, this allows us to propose the model sketched in the inset of Figure~\ref{fig6} (a). X$_1$ and X$_2$ are due to the recombination of two distinct excitons whose population depends on the temperature. Thermal repopulation of X$_1$ from X$_2$ (with an activation energy of 36.7\,meV in case of PEA$_2$SnI$_4$ (Fig.~S12), occurs at elevated temperature, but is suppressed upon cooling. This generates the rapid initial decay of X$_1$. At the same time, the slowed down transfer towards X$_2$, as indicated by the increasingly delayed onset, suggests that X$_2$ emission is preceded by carrier diffusion. In other words, the lower lying electronic state responsible for X$_2$ likely results from imperfections in the material.\\

We note that more recently, PL peaks below X$_1$ have become a hot topic with explanations invoking a variety of mechanisms including defects at grain boundaries,\cite{Wang2020a} a magnetic dipole transition,\cite{DeCrescent2020,DeCrescent2020a} and the formation of distinct excitons due to interaction with phonons.\cite{Li2020e,Moral2020} Several reports refer to the strong dependence of the detection direction and polarisation conditions relative to the lattice planes, with the lower transition particularly prominent for in-plane detection.\cite{Li2020e,Wang2020a,DeCrescent2020} In principle, our data are congruent with the proposition of defects at grain boundaries, as put forward by Wang \textit{et al.},\cite{Wang2020a} but since X$_2$ is similarly observed for single crystals, grain boundaries alone appear to be an unlikely origin. In our view, propositions that invoke polaronic effects, such as self-trapping, also fail to explain the corresponding absorption, the strong non-linear and partly irreversible effect of laser power, as well as the delayed maximum in the transient PL.\\
Taken together, the above evidence makes shallow defects and bound excitons strong contenders for the origin of X$_2$: although detectable, its absorption is much lower than the 1$s$ transition associated with X$_1$. Furthermore, it coincidences energetically with shallow defects in PbI$_2$ (Fig.~S6) and bound excitons in bulk perovskites\cite{Francisco-Lopez2021} and exhibits an activation energy similar to one attributed to shallow traps in diffusion experiments on PEA$_2$PbI$_4$ single crystals.\cite{Seitz2021} Non-linearities of the PL intensity could easily be explained through the well-known photoactivation of halide-related defects.\cite{DeQuilettes2016,Mosconi2016,Liu2016,Jiang2021} In this model, the strong polarisation dependence could be related to a distinct type of crystal imperfection.

\begin{figure}[]
	\centering
	\includegraphics[width=1\linewidth]{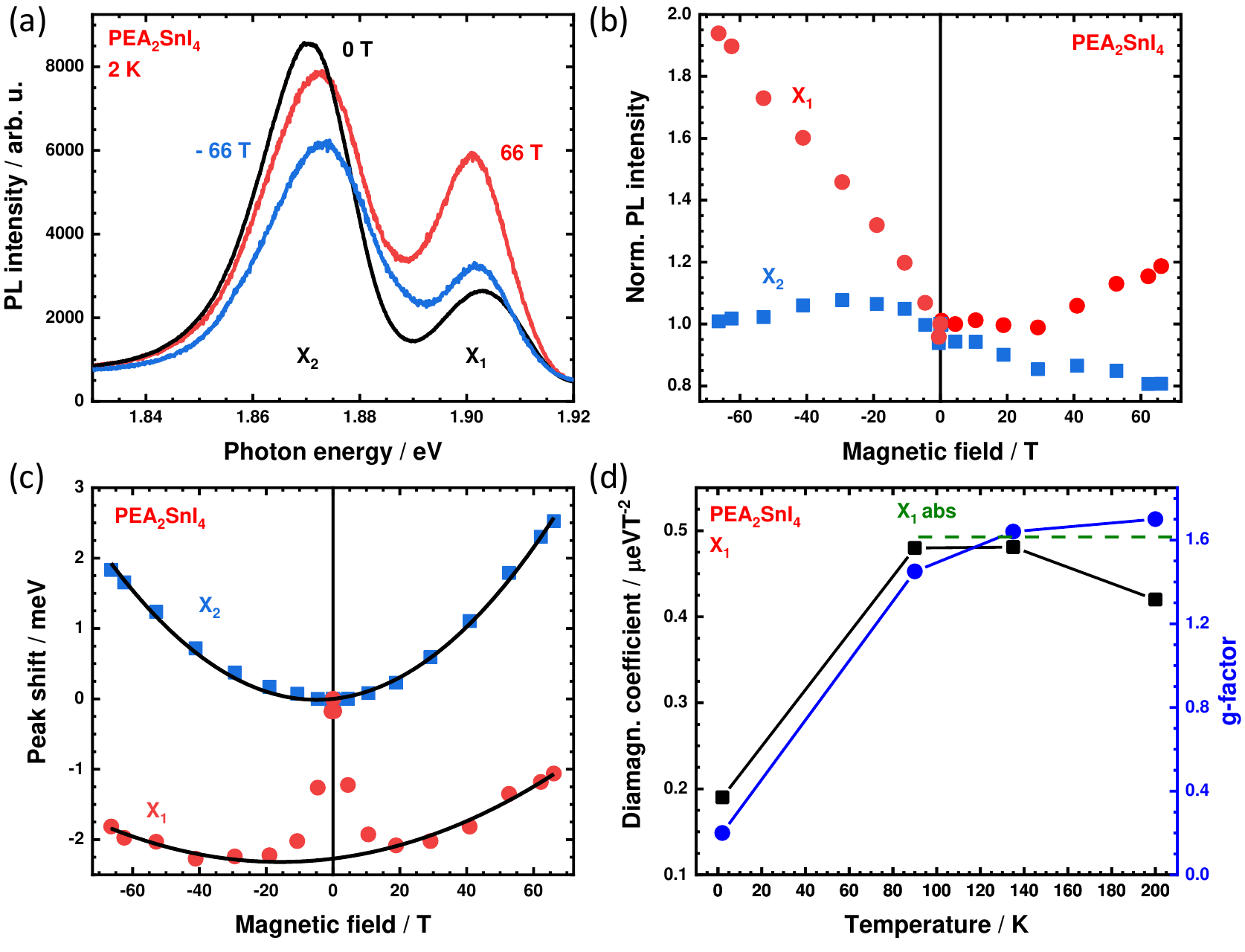}
	\caption{Magneto-photoluminescence data of PEA$_2$SnI$_4$ at 2\,K. (a) Absolute PL spectra under 66\,T field of opposite polarities (red and blue lines) and in zero field condition. (b) Integrated PL intensity of X$_1$ and X$_2$ relative to zero field condition as a function of the magnetic field. (c) Extracted peak position; black lines indicate fits to equation~1. Temperature dependence of the diamagnetic coefficient (black) and g-factor (blue) of X$_1$. The dashed red line indicates the g-factor of the 1\textit{s} exciton absorption at 2\,K.}
	\label{fig7}
\end{figure}

One of the remaining observations to account for is the prominent red-shift of X$_1$ at early delay times (Fig.~\ref{fig5} (g), (h)) observed specifically for the iodide-based variants (highlighted also in Fig.~S15-S17). Given the small Stokes shift of X$_1$, one might think the spectral shift to be due to increased reabsorption upon carrier diffusion into the bulk of the film, but as magneto-PL studies reveal, what we termed X$_1$ is actually the convolution of at least two emissive states.\\
Figure~\ref{fig7} (a) shows the PL of PEA$_2$SnI$_4$ when applying a magnetic field of 66\,T in out-of-plane direction (Faraday configuration) at 2\,K along with the spectrum under zero field condition. We probed the emission of $\sigma^+$ and $\sigma^-$ circularly polarized light (red, blue) and find the X$_1$ transition to increase in intensity for both field directions (Fig.~\ref{fig7} (a), (b)), whereas X$_2$ decreases slightly under high field strengths. Additional to the changes in intensity, the two main peaks also undergo a field-dependent energy shift, illustrated in Figure~\ref{fig7} (c). The black lines indicate fits to the interplay of Zeeman splitting and diamagnetic shift, as discussed below, but the low-field behaviour of X$_1$ shows an uncharacteristic red-shift of approximately 2\,meV.\\
Whereas brightening is generally attributed to either an increased oscillator strength imposed by a B-field induced reduction of the exciton's spatial extent or by the intermixing of bright and dark states,
\cite{Miura2002,Neumann2021,Miura2007,Zhang2015h} the low-field red-shift of X$_1$ clearly points to the presence of a lower lying state that brightens at higher field strengths. Such a field-induced brightening of the lower lying state is more clearly observed in the case of PEA$_2$PbI$_4$ (Supplementary Note~6.)

Additional insight can be obtained when considering how the diamagnetic shift and the Zeeman effect impact the luminescence. The magnitude of the exciton energy shift in the low-field limit is given by:\cite{Dyksik2020}
\begin{equation}
	\Delta E_\pm=\pm1/2 g_{eff} \mu_B B+c_0 B^2	,
\end{equation}
in which $g_{eff}$ denotes the effective Land\'{e} g-factor, $\mu_B$ the Bohr magneton and $c_0$ the diamagnetic coefficient. Fitting equation (1) to the data in Figure\ref{fig7} (c), we find a diamagnetic shift of 0.19 and a g-factor of 0.20 for X$_1$, when excluding the low-field shift (for X$_2$ the values are 0.51\,$\mu$eVT$^{-2}$ and 0.17). The diamagnetic shift as well as the g-factor are considerably lower than the reported values for the free exciton absorption in this material: $c_0$=0.68 $\mu$eVT$^{-2}$ and $g_{eff}$=1.6.\cite{Dyksik2020} This indicates that low temperature emission occurs from considerably more localised states than the absorbance. When the temperature is raised, $c_0$ and  $g_{eff}$ of the emission increase (Fig.~\ref{fig7} (d)) to approach the values corresponding to the free exciton absorption at 2\,K (indicated by the dashed red line for $g_{eff}$).\\
The picture that emerges for PEA$_2$SnI$_4$ is thus as follows: in accordance with the time resolved data, the X$_1$ emission stems from two states at low temperature, of which the lower one is localised, likely a bound exciton, and becomes increasingly bright under the influence of an external magnetic field. Temperature increase lifts the localisation and the overall emission becomes governed by the same transition responsible for photoabsorption, that is the free 1$s$ exciton.

The PEA$_2$PbI$_4$ data are largely analogous to the tin-based variant and we discuss the results in Supplemetary Note~6 (Fig.~S21, S22). It is noteworthy that for PEA$_2$PbI$_4$ the lower-lying state of X$_1$ can be linked to the defect induced emission identified, for example through the strong and inverse energy shift upon temperature variation (Fig.~S21). This contrasts with the case of PEA$_2$SnI$_4$, where no distinct peak can be observed between X$_1$ and X$_2$ at any temperature. These findings for the lead-based variant support the localised nature of the low-lying state.\\
Splitting of X$_1$ was recently also reported for single crystals of PEA$_2$PbI$_4$ based on a similar experimental approach,\cite{Do2020} where the peaks were said to be due to two distinct bright states. Alternatively, theoretical considerations and transient PL studies have led to the proposition of an unresolved pair of low-lying dark and higher-lying bright state to form X$_1$ in PEA$_2$PbI$_4$.\cite{Fang2020} This line of reasoning followed earlier studies on bromide-based compounds\cite{Tanaka2005} and was also invoked to explain aspects of the temperature dependent PL of PEA$_2$SnI$_4$.\cite{Folpini2020}\\
These reports, however, base their explanations on properties of the free exciton and do not consider additional extrinsic effects, as suggested by the localisation. Splitting into a low-lying dark state and a higher bright state, in particular, is a compelling explanation for the long PL lifetime of X$_1$ and the B-field induced brightening. A recent report on PEA$_2$PbI$_4$ has consequently ascribed this dark nature to the localised state at 2.32\,eV.\cite{Neumann2021} However, as the original dark-bright splitting was said to be due to the electron-hole exchange interaction of the free exciton,\cite{Tanaka2005,Fang2020} the underlying mechanism for this attribution is unclear. Either way, our magneto-PL data show that a mechanism invoked to explain the behaviour of PEA$_2$SnI$_4$ will also have to apply to PEA$_2$PbI$_4$ and we note that strong evidence for the impact of carrier localisation on the luminescence has also been found in studies on carrier diffusion in these materials.\cite{Ziegler2020}

\textbf{CONCLUSION}

In summary, we studied the photophysics of two-dimensional metal halide perovskites with varied composition of the inorganic constituents. The recombination of free excitons generates a single pronounced emission line at room temperature, which narrows upon sample cooling to reveal an intricate substructure. We attribute high energy features to a vibronic progression from transitions that break Kasha's rule.\\
Our study highlights a lack in the current understanding of the vibronic progression, as the spacing of the high energy PL lines cannot be traced back to phonon modes observable in Raman spectroscopy and cannot be linked directly to the organic cation on the A-site. Our findings not only warrant further study on the excited state vibrational properties, but also highlight an unexpectedly long-lived tail of the hot emission, whose origin is as yet unclear.\\
All samples exhibit two main emission lines, of which the upper one is attributed to the 1$s$ exciton. By means of time resolved luminescence and magneto-PL studies, we show that this bright peak generally attributed to free exciton recombination is split into two transitions, of which the lower is due to a localised state that shows evidence of being optically dark. Based on temperature and time dependent studies the second bright emission peak is linked to crystal imperfections that generate shallow defects.\\
Our work underlines the largely similar photophysics of these two-dimensional compounds and shows how metal-halide substitution affects their photophysics. Close study not only reveal important insights, but highlights and uncovers open questions.

\textbf{Acknowledgement}

	Arjen Kamp and Teodor Zaharia are thanked for technical support. We thank Dr. Graeme Blake for insightful discussions on the crystallography. SK is grateful for a research fellowship (Grant No: 408012143) awarded by the Deutsche Forschungsgemeinschaft (DFG). This work was financed through the Materials for Sustainability (Mat4Sus) programme (739.017.005) of the Netherlands Organisation for Scientific Research (NWO). This work is part of the research program of the Foundation for Fundamental Research on Matter (FOM), which is part of NWO. This is a publication of the FOM‐focus Group "Next Generation Organic Photovoltaics" participating in the Dutch Institute for Fundamental Energy Research (DIFFER). The project received funding from National Science Centre Poland within the OPUS program (grant no. 2019/33/B/ST3/01915). This work was partially supported by the OPEP project, which received funding from the ANR-10-LABX-0037-NEXT. M.D. appreciates support from the Polish National Agency for Academic Exchange within the Bekker programme (grant no. PPN/BEK/2019/1/00312/U/00001). The Polish participation in the European Magnetic Field Laboratory is supported by the DIR/WK/2018/07 grant from Ministry of Science and Higher Education, Poland.

\bibliographystyle{advfunctmater}
\bibliography{library}

\begin{thebibliography}{10}

\bibitem{Ishihara1989}
T.~Ishihara, J.~Takahashi, T.~Goto.
\newblock \textit{Solid State Commun.} \textit{69}, 933, \textbf{1989}.

\bibitem{Blancon2020}
J.-C. Blancon, J.~Even, C.~C. Stoumpos, M.~G. Kanatzidis, A.~D. Mohite.
\newblock \textit{Nat. Nanotechnol.} \textit{15}, 969, \textbf{2020}.

\bibitem{Marchenko2020}
E.~I. Marchenko, S.~A. Fateev, A.~A. Petrov, V.~V. Korolev, A.~Mitrofanov,
  A.~V. Petrov, E.~A. Goodilin, A.~B. Tarasov.
\newblock \textit{Chem. Mater.} \textit{32}, 7383, \textbf{2020}.

\bibitem{Du2017a}
K.-Z. Du, Q.~Tu, X.~Zhang, Q.~Han, J.~Liu, S.~Zauscher, D.~B. Mitzi.
\newblock \textit{Inorg. Chem.} \textit{56}, 9291, \textbf{2017}.

\bibitem{Katan2019}
C.~Katan, N.~Mercier, J.~Even.
\newblock \textit{Chem. Rev.} \textit{119}, 3140, \textbf{2019}.

\bibitem{Dyksik2020}
M.~Dyksik, H.~Duim, X.~Zhu, Z.~Yang, M.~Gen, Y.~Kohama, S.~Adjokatse, D.~K.
  Maude, M.~A. Loi, D.~A. Egger, M.~Baranowski, P.~Plochocka.
\newblock \textit{ACS Energy Lett.} \textit{5}, 3609, \textbf{2020}.

\bibitem{Fang2020}
H.~H. Fang, J.~Yang, S.~Adjokatse, E.~Tekelenburg, M.~E. Kamminga, H.~Duim,
  J.~Ye, G.~R. Blake, J.~Even, M.~A. Loi.
\newblock \textit{Adv. Funct. Mater.} \textit{30}, 1907979, \textbf{2020}.

\bibitem{Do2020}
T.~T.~H. Do, A.~{Granados del {\'{A}}guila}, D.~Zhang, J.~Xing, S.~Liu, M.~A.
  Prosnikov, W.~Gao, K.~Chang, P.~C.~M. Christianen, Q.~Xiong.
\newblock \textit{Nano Lett.} \textbf{2020}.

\bibitem{Kahmann2020}
S.~Kahmann, E.~K. Tekelenburg, H.~Duim, M.~E. Kamminga, M.~A. Loi.
\newblock \textit{Nat. Commun.} \textit{11}, 2344, \textbf{2020}.

\bibitem{Straus2016}
D.~B. Straus, S.~{Hurtado Parra}, N.~Iotov, J.~Gebhardt, A.~M. Rappe, J.~E.
  Subotnik, J.~M. Kikkawa, C.~R. Kagan.
\newblock \textit{J. Am. Chem. Soc.} \textit{138}, 13798, \textbf{2016}.

\bibitem{Thouin2018}
A.~R. {Srimath Kandada}, D.~Beljonne, C.~Silva, A.~Petrozza, F.~Thouin, D.~A.
  Valverde-Ch{\'{a}}vez, C.~Quarti, D.~Cortecchia, I.~Bargigia.
\newblock \textit{Nat. Mater.} \textbf{2018}.

\bibitem{Straus2019}
D.~B. Straus, N.~Iotov, M.~R. Gau, Q.~Zhao, P.~J. Carroll, C.~R. Kagan.
\newblock \textit{J. Phys. Chem. Lett.} \textit{10}, 1198, \textbf{2019}.

\bibitem{Straus2020}
D.~B. Straus, S.~{Hurtado Parra}, N.~Iotov, Q.~Zhao, M.~R. Gau, P.~J. Carroll,
  J.~M. Kikkawa, C.~R. Kagan.
\newblock \textit{ACS Nano} \textit{14}, 3621, \textbf{2020}.

\bibitem{Folpini2020}
G.~Folpini, D.~Cortecchia, A.~Petrozza, A.~R. {Srimath Kandada}.
\newblock \textit{J. Mater. Chem. C} \textit{8}, 10889, \textbf{2020}.

\bibitem{Tekelenburg2021}
E.~K. Tekelenburg, S.~Kahmann, M.~E. Kamminga, G.~R. Blake, M.~A. Loi.
\newblock \textit{Adv. Opt. Mater.} page 2001647, \textbf{2021}.

\bibitem{Fabini2020}
D.~H. Fabini, R.~Seshadri, M.~G. Kanatzidis.
\newblock \textit{MRS Bull.} \textit{45}, 467, \textbf{2020}.

\bibitem{Zhang2020a}
L.~Zhang, X.~Zhang, G.~Lu.
\newblock \textit{J. Phys. Chem. Lett.} \textit{11}, 6982, \textbf{2020}.

\bibitem{Goesten2018}
M.~G. Goesten, R.~Hoffmann.
\newblock \textit{J. Am. Chem. Soc.} \textit{140}, 12996, \textbf{2018}.

\bibitem{Parrott2018}
E.~S. Parrott, T.~Green, R.~L. Milot, M.~B. Johnston, H.~J. Snaith, L.~M. Herz.
\newblock \textit{Adv. Funct. Mater.} \textit{28}, 1802803, \textbf{2018}.

\bibitem{Kahmann2019a}
S.~Kahmann, S.~Shao, M.~A. Loi.
\newblock \textit{Adv. Funct. Mater.} \textit{29}, 1902963, \textbf{2019}.

\bibitem{Urban2020}
J.~M. Urban, G.~Chehade, M.~Dyksik, M.~Menahem, A.~Surrente,
  G.~Tripp{\'{e}}-Allard, D.~K. Maude, D.~Garrot, O.~Yaffe, E.~Deleporte,
  P.~Plochocka, M.~Baranowski.
\newblock \textit{J. Phys. Chem. Lett.} \textit{11}, 5830, \textbf{2020}.

\bibitem{Niemann2016}
R.~G. Niemann, A.~G. Kontos, D.~Palles, E.~I. Kamitsos, A.~Kaltzoglou,
  F.~Brivio, P.~Falaras, P.~J. Cameron.
\newblock \textit{J. Phys. Chem. C} \textit{120}, 2509, \textbf{2016}.

\bibitem{Dhanabalan2020}
B.~Dhanabalan, Y.-C. Leng, G.~Biffi, M.-L. Lin, P.-H. Tan, I.~Infante,
  L.~Manna, M.~P. Arciniegas, R.~Krahne.
\newblock \textit{ACS Nano} \textit{14}, 4689, \textbf{2020}.

\bibitem{Uchinokura1974}
K.~Uchinokura, T.~Sekine, E.~Matsuura.
\newblock \textit{J. Phys. Chem. Solids} \textit{35}, 171, \textbf{1974}.

\bibitem{Ni2017a}
L.~Ni, U.~Huynh, A.~Cheminal, T.~H. Thomas, R.~Shivanna, T.~F. Hinrichsen,
  S.~Ahmad, A.~Sadhanala, A.~Rao.
\newblock \textit{ACS Nano} \textit{11}, 10834, \textbf{2017}.

\bibitem{Kato2003}
Y.~Kato, D.~Ichii, K.~Ohashi, H.~Kunugita, K.~Ema, K.~Tanaka, T.~Takahashi,
  T.~Kondo.
\newblock \textit{Solid State Commun.} \textit{128}, 15, \textbf{2003}.

\bibitem{Tanaka2005}
K.~Tanaka, T.~Takahashi, T.~Kondo, K.~Umeda, K.~Ema, T.~Umebayashi, K.~Asai,
  K.~Uchida, N.~Miura.
\newblock \textit{Jpn. J. Appl. Phys.} \textit{44}, 5923, \textbf{2005}.

\bibitem{Goto2006}
T.~Goto, H.~Makino, T.~Yao, C.~H. Chia, T.~Makino, Y.~Segawa, G.~A. Mousdis,
  G.~C. Papavassiliou.
\newblock \textit{Phys. Rev. B} \textit{73}, 115206, \textbf{2006}.

\bibitem{Neutzner2018}
S.~Neutzner, F.~Thouin, D.~Cortecchia, A.~Petrozza, C.~Silva, A.~R. {Srimath
  Kandada}.
\newblock \textit{Phys. Rev. Mater.} \textit{2}, 064605, \textbf{2018}.

\bibitem{Li2019h}
J.~Li, J.~Wang, J.~Ma, H.~Shen, L.~Li, X.~Duan, D.~Li.
\newblock \textit{Nat. Commun.} \textit{10}, 806, \textbf{2019}.

\bibitem{Wang2020a}
M.~Wang, J.~Tang, H.~Wang, C.~Zhang, Y.~S. Zhao, J.~Yao.
\newblock \textit{Adv. Opt. Mater.} \textit{8}, 1901780, \textbf{2020}.

\bibitem{DeCrescent2020}
R.~A. DeCrescent, X.~Du, R.~M. Kennard, N.~R. Venkatesan, C.~J. Dahlman, M.~L.
  Chabinyc, J.~A. Schuller.
\newblock \textit{ACS Nano} \textit{14}, 8958, \textbf{2020}.

\bibitem{DeCrescent2020a}
R.~A. DeCrescent, N.~R. Venkatesan, C.~J. Dahlman, R.~M. Kennard, X.~Zhang,
  W.~Li, X.~Du, M.~L. Chabinyc, R.~Zia, J.~A. Schuller.
\newblock \textit{Sci. Adv.} \textit{6}, eaay4900, \textbf{2020}.

\bibitem{Li2020e}
J.~Li, J.~Ma, X.~Cheng, Z.~Liu, Y.~Chen, D.~Li.
\newblock \textit{ACS Nano} \textit{14}, 2156, \textbf{2020}.

\bibitem{Moral2020}
R.~F. Moral, J.~C. Germino, L.~G. Bonato, D.~B. Almeida, E.~M. Ther{\'{e}}zio,
  T.~D.~Z. Atvars, S.~D. Stranks, R.~A. Nome, A.~F. Nogueira.
\newblock \textit{Adv. Opt. Mater.} \textit{2001431}, 2001431, \textbf{2020}.

\bibitem{Francisco-Lopez2021}
A.~Francisco‐L{\'{o}}pez, B.~Charles, M.~I. Alonso, M.~Garriga, M.~T. Weller,
  A.~R. Go{\~{n}}i.
\newblock \textit{Adv. Opt. Mater.} \textit{2001969}, 2001969, \textbf{2021}.

\bibitem{Seitz2021}
M.~Seitz, M.~Mel{\'{e}}ndez, N.~Alc{\'{a}}zar‐Cano, D.~N. Congreve,
  R.~Delgado‐Buscalioni, F.~Prins.
\newblock \textit{Adv. Opt. Mater.} \textit{2001875}, 2001875, \textbf{2021}.

\bibitem{DeQuilettes2016}
D.~W. DeQuilettes, W.~Zhang, V.~M. Burlakov, D.~J. Graham, T.~Leijtens,
  A.~Osherov, V.~Bulovi{\'{c}}, H.~J. Snaith, D.~S. Ginger, S.~D. Stranks.
\newblock \textit{Nat. Commun.} \textit{7}, 11683, \textbf{2016}.

\bibitem{Mosconi2016}
E.~Mosconi, D.~Meggiolaro, H.~J. Snaith, S.~D. Stranks, F.~{De Angelis}.
\newblock \textit{Energy Environ. Sci.} \textit{9}, 3180, \textbf{2016}.

\bibitem{Liu2016}
M.~Liu, O.~Voznyy, R.~Sabatini, F.~P. {Garc{\'{i}}a de Arquer}, R.~Munir, A.~H.
  Balawi, X.~Lan, F.~Fan, G.~Walters, A.~R. Kirmani, S.~Hoogland, F.~Laquai,
  A.~Amassian, E.~H. Sargent.
\newblock \textit{Nat. Mater.} \textit{1}, 1, \textbf{2016}.

\bibitem{Jiang2021}
F.~Jiang, J.~Pothoof, F.~Muckel, R.~Giridharagopal, J.~Wang, D.~S. Ginger.
\newblock \textit{ACS Energy Lett.} \textit{6}, 100, \textbf{2021}.

\bibitem{Miura2002}
N.~Miura, K.~Uchida, T.~Yasuhira, E.~Kurtz, C.~Klingshirn, H.~Nakashima,
  F.~Issiki, Y.~Shiraki.
\newblock \textit{Phys. E Low-dimensional Syst. Nanostructures} \textit{13},
  263, \textbf{2002}.

\bibitem{Neumann2021}
T.~Neumann, S.~Feldmann, P.~Moser, J.~Zerhoch, T.~van~de Goor, A.~Delhomme,
  T.~Winkler, J.~J. Finley, C.~Faugeras, M.~S. Brandt, A.~V. Stier,
  F.~Deschler.
\newblock \textit{arXiv} \textbf{2020}.

\bibitem{Miura2007}
N.~Miura.
\newblock \textit{{Physics of Semiconductors in High Magnetic Fields}}.
\newblock Oxford University Press, \textbf{2007}.

\bibitem{Zhang2015h}
C.~Zhang, D.~Sun, C.-X. Sheng, Y.~X. Zhai, K.~Mielczarek, A.~Zakhidov, Z.~V.
  Vardeny.
\newblock \textit{Nat. Phys.} \textit{11}, 427, \textbf{2015}.

\bibitem{Ziegler2020}
J.~D. Ziegler, J.~Zipfel, B.~Meisinger, M.~Menahem, X.~Zhu, T.~Taniguchi,
  K.~Watanabe, O.~Yaffe, D.~A. Egger, A.~Chernikov.
\newblock \textit{Nano Lett.} \textit{20}, 6674, \textbf{2020}.

\end{thebibliography}

\end{document}